\documentclass{elsart}
\include{psfig}
\begin{document}
\begin{frontmatter}
\title{Effect of the intermediate velocity emissions
on the quasi-projectile properties for the 
Ar+Ni system at 95 A.MeV }
\author[Saclay,Orsay]{D.~Dor\'e}\footnote{DAPNIA/SPhN, CEA/Saclay,
91191 Gif-sur-Yvette Cedex, France, e-mail: dore@in2p3.fr, FAX : 33.1.69.08.75.84 },
\author[Saclay]{Ph.~Buchet},
\author[Saclay]{J.L.~Charvet},
\author[Saclay]{R.~Dayras},
\author[Saclay]{L.~Nalpas},
\author[LPC]{D.~Cussol},
\author[LPC]{T.~Lefort},
\author[Saclay]{R.~Legrain},
\author[Saclay]{C.~Volant},
\author[GANIL]{G.~Auger},
\author[Orsay]{Ch.O.~Bacri},
\author[LPC]{N.~Bellaize},
\author[LPC]{F.~Bocage},
\author[LPC]{R.~Bougault},
\author[GANIL]{B.~Bouriquet},
\author[LPC]{R.~Brou},
\author[GANIL]{A.~Chbihi},
\author[LPC]{J.~Colin},
\author[lyon]{A.~Demeyer},
\author[LPC]{D.~Durand},
\author[GANIL]{J.D.~Frankland},
\author[Orsay,cnam]{E.~Galichet},
\author[LPC]{E.~Genouin-Duhamel},
\author[lyon]{E.~Gerlic},
\author[lyon]{D.~Guinet},
\author[GANIL]{S.~Hudan},
\author[lyon]{P.~Lautesse},
\author[Orsay]{F.~Lavaud},
\author[GANIL]{J.L.~Laville},
\author[LPC]{J.F.~Lecolley},
\author[lyon]{C.~Leduc},
\author[LPC]{N.~Le Neindre},
\author[LPC]{O.~Lopez},
\author[LPC]{M.~Louvel},
\author[lyon]{A.M.~Maskay},
\author[LPC]{J.~Normand},
\author[roum]{M.~Parlog},
\author[Orsay]{P.~Pawlowski},
\author[Orsay]{E.~Plagnol},
\author[Orsay]{M.F.~Rivet},
\author[ita]{E.~Rosato},
\author[GANIL]{F.~Saint-Laurent}\footnote{present address : 
CEA, DRFC/STEP, 
CE Cadarache, 13108 Saint-Paul-lez-Durance, France},
\author[LPC]{J.C.~Steckmeyer},
\author[lyon]{M.~Stern},
\author[Orsay]{G.~Tabacaru},
\author[LPC]{B.~Tamain},
\author[Orsay]{L.~Tassan-Got},
\author[GANIL]{O.~Tirel},
\author[LPC]{E.~Vient},
\author[GANIL]{J.P.~Wieleczko}
\collab [INDRA Collaboration]
\address[Saclay]{ DAPNIA/SPhN, CEA/Saclay,
91191 Gif-sur-Yvette Cedex, France}
\address[LPC]{LPC Caen
(IN2P3-CNRS/ISMRA et Universit\'e),
14050 Caen Cedex , France}
\address[GANIL]{GANIL (DSM-CEA/IN2P3-CNRS),
B.P. 5027, 14076 Caen Cedex 5, France}
\address[Orsay]{IPN Orsay (IN2P3-CNRS),
91406 Orsay Cedex, France}
\address[lyon]{IPN Lyon (IN2P3-CNRS/Universit\'e),
69622 Villeurbanne Cedex, France}
\address[roum]{Nuclear Institute for Physics and Nuclear Engineering, 
Bucharest, Romania}
\address[ita]{Dipartimento di Scienze Fisiche, Univ. di Napoli, 180126 
Napoli, Italy}
\address[cnam]{Conservatoire National des Arts et M\'etiers, 75141 
Paris Cedex 03, France.}
\newpage
\begin{abstract}
The quasi-projectile (QP) properties are investigated in the Ar+Ni 
collisions at 95 A.MeV taking into account the intermediate velocity 
emission. Indeed, in this reaction, between 52 and 95 A.MeV bombarding 
energies, the number of particles emitted in the intermediate velocity 
region is related to the overlap volume between projectile and target. 
Mean transverse energies of these particles are found particularly high.
In this context, the mass of the QP decreases linearly with the impact 
parameter from peripheral to central collisions whereas its excitation 
energy increases up to 8 A.MeV. These results are compared to previous 
analyses assuming a pure binary scenario.
\end{abstract}
\end{frontmatter}

The Quasi-Projectile deexcitation has been studied through a wide 
variety of systems at intermediate energies \cite{guerreau}-
\cite{angel}.  In this energy domain a transition from a binary process,
leading to two main excited fragments (the quasi-projectile (QP) and 
the quasi-target (QT)) in the exit channel, towards a 
participant-spectator mechanism, is expected. From inclusive or semi 
exclusive 
measurements it has not been possible to distinguish between these two
mechanisms. In some cases, the experimental data could be described 
equally well either assuming a pure binary mechanism or a geometrical 
process \cite{roland},\cite{Morissey}. With the improvement of 
experimental setups, namely the advent of 4$\pi$ multidetectors 
allowing fully exclusive measurements, it should become possible to 
reconstruct the  QP and the QT from their decay products on an event
by event basis.  However, this reconstruction process depends greatly 
on  our ability to identify unambiguously the origin of the detected
products. Unfortunately, in the intermediate energy range, the various
sources of emission strongly overlap in the velocity space. Thus one 
has to rely on some assumptions on the underlying mechanisms in order
to unfold the various sources of emission.  Although it was generally 
admitted that below 100 AMeV, heavy ion collisions have essentially a 
binary character, it has been shown since several years that the decay
products could not be fully imputed to the decay of excited quasi-
projectile and quasi-target \cite{peter}-\cite{angel},\cite{montoya}-
\cite{tho}.  Besides preequilibrium and direct emissions already 
observed at low energies, processes like neck emission and aligned 
fission had to be taken into account in order to explain the 
experimental data. Indeed an excess of particles and fragments, not 
explained by the statistical deexcitation of fully equilibrated
QP and QT, is observed at intermediate velocity with unusual kinematical
properties.

From recent experimental data on the Ar+Ni reactions between 52 and 95 
A.MeV obtained at GANIL with the 4$\pi$ multidetector INDRA it was shown 
\cite{tho} that it was not possible to reproduce the light particle 
rapidity spectra by assuming only statistical emissions from excited QP 
and QT and that there was an excess of high energy particles at mid-
rapidity which increases as the impact parameter decreases. In the 
present paper, we will concentrate on the Ar+Ni reaction at 95 AMeV and
we will show how the properties of the QP that one can extract from the data
are strongly affected by particle emission around mid-rapidity. First 
the impact parameter classification and the event selection will be 
presented. Then, the QP properties, mass and excitation energy, will be 
established according to two basic assumptions. i) Neglecting mid-
rapidity emission, following previous analysis \cite{vap}-\cite{calo}, 
a two source reconstruction will be performed in the frame of a purely 
binary scenario. ii) In an attempt to take into account mid-rapidity 
emission (MRE) as evidenced in \cite{tho} we will unfold the experimental 
light charged particle rapidity spectra assuming three sources of 
emission, the QP, the QT and a third source of emission to simulate
the mid-rapidity contribution.  Thermal and shape equilibrium are assumed
in each source.  Then the properties of the QP are extracted from its 
decay products.  In both cases, the mass and the excitation energy of 
the QP thus obtained will be presented as a function of an experimental 
impact parameter. Finally, results of both reconstruction methods will 
be compared and discussed.\\

The experiment was performed at the GANIL facility which provided an
$^{36}$Ar beam of 3-4 x 10$^7$ pps at 95 A.MeV.  After collision with 
a 193 $\mu$g/cm$^2$ self-supporting $^{58}$Ni target, reaction products 
were detected with the 4$\pi$ charged particle detector INDRA \cite{pou} 
with a minimum bias trigger requiring a four fold event. Charge identification
is achieved up to the projectile charge in the forward hemisphere. 
Hydrogen and helium isotopes are separated for detection angles from 
3$^{0}$ to 176$^{0}$ (rings 2 to 17).  

Using the prescription of ref. \cite{cavata}, an impact parameter scale 
($b_{exp}$) is deduced from the total transverse energy distribution 
($E_{tr}^{tot}$) for all detected events as shown by the full line in Fig. 1(a).  
  For the 
forthcoming analysis, we will retain only events for which, at 
least the remnant of the QP has been detected. This is done using the 
correlation between the total detected charge ($Z_{tot}$) and the pseudo 
total parallel momentum ($P_{// tot}=\Sigma Z_i \times V_{// i}$)
 presented in Fig. 1(b). Only events for which 
$P_{// tot} \geq 70\%P_{proj}$ are kept. This condition selects events
with $Z_{tot}$ around and larger than the projectile charge and 
represents $\approx 60\%$ of the estimated \cite{kox} total reaction 
cross section, $\sigma^{th}_r$, whereas the total detected one amounts to 
80$\%$ of $\sigma^{th}_r$.  We remark that the selected events (dashed line in Fig. 
1(a)) still cover the whole range of $E^{tot}_{tr}$.\\

Proton and alpha particle reduced rapidity spectra (Y/Y$_p$ where Y$_p$ 
is the projectile rapidity) show two components centered respectively 
around the target and the projectile velocities as shown in Fig. 2  
for protons at $b_{exp}$=6 fm. This strongly suggests evaporation from 
excited QP and QT. Then, assuming a binary scenario and neglecting any 
non equilibrated emissions \cite{vap},\cite{metivier}, all particles 
and fragments are attributed to the QP or the QT event-by-event. The 
reconstruction is based on a simplified version of the thrust method 
\cite{thrust}. Both procedures roughly allocate all particles and 
fragments with a parallel velocity smaller than the center-of-mass 
velocity to the QT and the others to the QP. Charges, masses and 
velocities of both sources are then calculated. Neutrons added 
in order to obtain the total mass of the system are distributed 
between the QP and the QT according to the N/Z ratio (=1.04) of 
the system. From simulations \cite{evapo}, the neutron kinetic energies
are evaluated as the mean kinetic energy of the protons minus 2 MeV to 
take into account the absence of Coulomb barrier. Calorimetry is then 
used to calculate the excitation energy ($E^*$) of the QP. Event by event
we have $E^*=\Sigma_i(m_ic^2 + E_i) - m_sc^2$, where
$m_i$ is the mass of each particle/fragment, $E_i$ their kinetic energy 
in the QP frame and $m_s$ the mass of the source. The mass of the QP 
thus reconstructed (around 34) is almost independent of the impact 
parameter. In contrast, the excitation energy per nucleon 
increases almost linearly with decreasing impact parameter to reach 
18 A.MeV for central collisions. This value is in agreement
with the one obtained in \cite{calo} for violent collisions. 
These results are shown by the full circles in Fig. 4 and will be further
discussed in connection with the results of the second assumption. \\

It was shown in \cite{tho} that isotropic evaporation from excited QP 
and QT was not sufficient to explain the measured light products 
($Z\leq6$) rapidity spectra. In particular, there is an excess of 
particles emitted around mid-rapidity which cannot be 
explained by a simple overlap of the QP and QT emission spheres. This 
mid-rapidity contribution increases with decreasing $b_{exp}$. 
Furthermore, the average transverse energy ($<E_{tr}>$) of these
particles (fig. 3(a) full circles) is much higher than expected from 
evaporation. The same behaviour is observed for all products of $Z\leq6$,  
suggesting that the excess of particles at intermediate velocity 
has peculiar kinematical properties. It has to be noted that due to
detection thresholds, the $<E_{tr}>$  values around the target rapidity 
are artificially increased. 
  
In order to take into account this intermediate velocity component, 
besides the two evaporating sources, emission from a third source 
around mid-rapidity was assumed.  A fit procedure, widely used to 
modelize differential cross sections \cite{phair,milkau,hsi}, 
is performed supposing three thermalized sources.
The laboratory energy spectra are fitted with the sum of three 
Maxwellian distributions assuming volume emission \cite{goldhaber}: 
\vskip -0.5cm

$$ d^2\sigma/dEd\Omega = \sum_{i=1,3} N_{i} 
\sqrt{E_{l}} \hskip 0.2cm exp[-(E_{l} + Es_{i} - 2 
\sqrt{(E_{l} Es_{i})} cos(\theta _{l}))/T_{i}] 
\hskip 1.cm (1)$$
$N_{i}$, $Es_{i}=\frac{1}{2}M_{part}V^2_{source}$, $T_{i}$ being adjustable 
parameters and $E_{l}$, $\theta_{l}$ and $M_{part}$ are respectively the energy,
the angle and the mass of the emitted particle in the laboratory.
The number of parameters (9) can be reduced 
with the following assumptions : 1) the temperatures of the QP and of 
the QT are the same \cite{ita}, 
2) assuming that non equilibrium particles are emitted symmetrically 
around $90^{0}$ in the 
center of mass, from momentum conservation, the QP and QT 
parallel velocities are linked through the 
relation $V_{QT}=(M_{proj}/M_{target})(V_{proj}-V_{QP}$). 

Due to statistics, only energy spectra of light particles (p,d,t,$^3$He,$^4$He) are fitted.  
For each particle type, the
detection energy thresholds are adjusted in order to have the same value for all rings
 (independently of the experimental thresholds which may fluctuate slightly from one
detector to the other). For a given impact parameter bin and a given particle type, all energy 
spectra from ring 2 to ring 17 are fitted simultaneously \cite{tho}. As ring 1 (2-3$^{0}$) does not provide 
isotopic separation, it is not included in the fit.
Thus, a set of parameters is obtained for each light particle type 
and each impact parameter bin. As shown in Fig. 3(b), the overall quality of the fits is quite good. 
Distribution irregularities are due to experimental biases.  Solid angles are different from one 
ring to the other. The average angle of a ring being used to calculate the rapidity, the distributions
are slightly distorted. For protons, 
at $b_{exp}$=3 fm, it is found that the mid-rapidity component contributes significantly to the total proton
rapidity distribution and covers the whole rapidity range. Direct emissions evaluated with intranuclear
cascade
calculations give similar results \cite{Piotr}-\cite{casca}. One notes also that the average
transverse energies, $<E_{tr}>$ as a function of $Y/Y_p$ are well reproduced 
(Fig. 3(a)). 

The fit parameters for protons and alpha particles evolve rather 
smoothly with $b_{exp}$ from central to peripheral collisions (see Table 1).   
The proton source reduced rapidities are rather constant from central to peripheral collisions 
for the QP (0.91$<$$Y_{QP}/Y_p$$<$0.93) and the mid-rapidity source (0.46$<$$Y_{MRE}/Y_p$$<$0.49). 
For other particles, 
both rapidities increase with impact parameter. The apparent temperature of the QP increases significantly from 
peripheral to central collisions (Table 1). At a given impact parameter, different particle  
types yield different temperatures in contradiction with the equilibrium hypothesis. In 
\cite{ita}, similar deviations were observed and their possible origin discussed. 
It has been shown \cite{buchet} that introducing nucleon-alpha scattering could improve
 significantly the fit for
the alpha particle rapidity spectra. Thus, nucleon-cluster collisions in the region of overlap between 
projectile and target may be in part responsible for the discrepancies between the temperature 
parameters and source rapidities obtained in our 
simple three source fits for different particle types.
For all particles, the apparent temperatures of the mid-rapidity source are large (Table 1) 
and increase strongly from peripheral to central collisions where they reach
$\simeq$25-30 MeV. 
These variations with impact parameter can be explained in part by the fact that the total 
transverse energy is used as impact parameter selector. In effect, for instance, the temperature parameter is linked to the mean transverse
energy through the relation $<E_{tr}> = T$ for volume emission, thus establishing a correlation between the impact 
parameter selection and the temperature (see below). 

The contribution of each source to the rapidity distribution depends upon the impact parameter and 
the particle type. The multiplicities of particles emitted by the QP and the QT follow the same 
evolution. The proton multiplicity for the QP (Table 1) stays constant around 1.5 from peripheral 
to mid-central collisions and then decreases to reach 0.5 in central collisions. For alpha 
particles, the multiplicity starts at a value of 0.5 in peripheral collisions to reach a maximum 
around 1.2 in mid-central collisions and then decreases to reach a value of 0.6 in central 
collisions. This behavior can be understood if the size of the source decreases with decreasing
impact parameter while the temperature increases.
For other particles the evolution is intermediate between that of protons and alpha 
particles. By contrast the multiplicity of particles emitted near mid-rapidity increases strongly 
as the impact parameter decreases whatever the particle type. This evolution suggests a geometrical
 effect as we will discuss later.

In order to evaluate the robustness of these results, several tests have been performed.
Using different prescriptions to fit the data, constraining some of the parameters, 
adding Coulomb barriers, assuming surface emission instead of volume 
emission, lead essentially to the same evolution of the parameters (velocities, temperatures and 
multiplicities) with impact parameter.  
Assuming a surface emission for QP and QT,  their source temperatures are found slightly lower but 
the fits are in poorer agreement with the experimental data. 
Using the heaviest fragment in the forward hemisphere, $Z_{max}$, as an indicator of the 
impact parameter (the closer to the charge of the projectile the fragment charge is, the larger
is the impact parameter) avoids the correlation between the temperature and the
impact parameter \cite{buchet}. This procedure yields lower QP and QT temperatures but does not affect the relative 
contributions of each source. In \cite{tho}, another global variable, related
to the dissipated energy in the forward hemisphere, was used to select events
according to the violence of the collision and emissions between 75$^o$ and 105$^o$
in the mid-rapidity frame were studied. In this case, the temperature parameters of the 
mid-rapidity evolved from 17 to 20 MeV for protons with the centrality of the
collision. The event selection
 and the angular cut explain
the differences between these results and those presented here.  However, we remark that these
values are located inside our limits (see Table 1).  

The next step is the reconstruction of the QP as a function of impact parameter. The mean
multiplicity and energy of each particle emitted by the QP at each impact 
parameter bin are used.  Because the projectile has N=Z, neutrons are added assuming 
that neutron multiplicity is equal to proton multiplicity ($<mult_n>=<mult_p>)$.
The contribution of fragments at mid-rapidity being small, those are shared between the QP and 
the QT as in the two source analysis previously described. 
For a given  impact parameter,
the mass of the QP is calculated as,

$$ <A_{QP}>=\Sigma_i <mult_i>\times A_i \, + \, 
\Sigma_f<mult_f>\times <A_f> \,$$
$$\hskip 7.cm  +\, <mult_n>\times A_n \hskip 2.cm (2)$$ 

and its excitation 
energy is estimated through calorimetry, 
\vskip -1.cm

$$ <E^*_{QP}>=\Sigma_i <mult_i>\times(m_ic^2 + \frac{3}{2} T_i) \, + \,
\Sigma_f<mult_f>\times(<m_f>c^2 + <E_f>)$$
$$\hskip 4.cm  +\, <mult_n>\times(m_nc^2 + \frac{3}{2} T_p) \, - \, <m_{QP}c^2> \hskip 1.cm (3)$$

where $i$, $n$, $f$ are the index for light charge particles, neutrons and fragments 
and QP refers to the emitting source. $A$'s are the atomic masses, 
$m$'s, the masses, $<E_f>$ are the fragment mean kinetic energies. Neutron temperature
($T_n$) is assumed to be equal to proton temperature ($T_p$).  One can note that 
fragments have an important contribution in (2) due to their masses and a small one in (3)
 due to their low kinetic energies. All this reconstruction assumes
  that particles originate from the same source even if the velocities
obtained with the fits are different.  This difference being larger for small impact 
parameter, values below 3 fm are less significant.

The QP masses and excitation energies thus obtained, are presented (stars) in Fig. (4) as 
a function of the impact parameter, together with the results (circles) of the previous 
two source analysis.  
Whereas a two source analysis yielded QP masses independent of impact 
parameter, in contrast, for the three source approach, one notes in 
fig. 4 (a) a linear increase with 
impact parameter of the QP mass : from 10 for central collisions to 32 for peripheral ones.
For peripheral collisions containing few mid-rapidity particles, both scenarii lead to nearly
identical results.
The linear mass increase with impact parameter suggests a  geometrical dependence. 
In fig. 4(a) the curve represents the QP mass predicted in
a calculation \cite{dayras} where the geometrical overlap of projectile and target 
is considered as the
intermediate source and the non interacting volumes are taken as QP and QT.
Although the general trend with impact parameter is similar to the three source 
result, 
the predicted QP mass decreases more rapidly with decreasing $b_{exp}$ than obtained from 
the three source analysis.  This discrepancy at low impact parameter may be imputed, in part,
to the $b_{exp}$ determination.  One can also argue that at 95 A.MeV the participant-spectator regime is
not fully reached.
 
Dynamical calculations for small systems \cite{peter},\cite{eudes2}-\cite{eudes3} present
 a similar 
relation between the QP mass and the impact parameter if particles emitted before the 
re-separation time of QP and QT are not included in the QP reconstruction.  These calculations also show
that these "early" particles are distributed over the whole range of parallel velocity
as deduced from the three source fits.  To obtain a realistic estimation of the QP emissions,
it is important to subtract the mid-rapidity component over the whole rapidity range.

Excitation energies deduced from both analyses are compared in fig. 4 (b).
The $Z_{max}$ sorting (open symbols) has also been tested in order to roughly evaluate the
 effect of 
the event sorting. The $Z_{max}$ value is the one corresponding to the more abundant
QP residue in the considered $b_{exp}$ bins.  As observed, the obtained values are
close to those of the three source fit method based on the $b_{exp}$ sorting, 
indicating that the sorting
has only small effect on the results.  In all cases the excitation energies increase with decreasing impact parameter. 
However, except for the most peripheral collisions where mid-rapidity emission is negligible, 
the three source fit method yields excitation energies about a factor of two smaller 
than the two source analysis.  As $b_{exp}$ decreases, the mid-rapidity component
carries an
increasing amount of the deposited energy, limiting the excitation energy imparted to the QP
and the QT.

Preliminary analyzes between 32 and 95 A.MeV \cite{bormio} show that beyond 52 A.MeV the
yields
of the different sources become independent of the bombarding energy.  The
mean transverse energy of the mid-rapidity component increases linearly with bombarding 
energy while it is constant for the QP and QT contributions. Above $\sim$ 50 
A.MeV, most of the energy is deposited into the overlap region between
projectile and target and is evacuated by the mid-rapidity particles.\\

Quasi-projectiles produced in the reaction $^{36}Ar+^{58}Ni$ at 95 A.MeV have been 
reconstructed from their decay products under two basic assumptions, $i)$ purely binary 
collisions, $ii)$ additional emission from the overlapping zone between projectile and target.
The properties (mass and excitation energy) of the QP thus reconstructed depend strongly upon 
these assumptions. Indeed, for mid-central collisions, there is a factor of 1.71 between QP masses
and 1.76 between excitation energies, the assumption $ii)$ leading to the lowest estimation.
It has been shown that the properties of particles emitted at mid-rapidity are incompatible 
with an evaporation process from fully equilibrated quasi-projectiles and quasi-targets which
made the assumption $i)$ unrealistic. Based upon results of $ii)$, these particles cover the 
whole rapidity range 
and mix in part with particles evaporated from the excited quasi-projectile and quasi-target.
The unfolding procedure presented in this work is an important step in order to reconstruct sources
with precision. 
The additional source in assumption $ii)$ includes many processes and it will be
necessary to disentangle them to go further in the interpretation.

\newpage
\begin{table} 
\begin{center}
\begin{tabular}{|c|c|c|c|c|c|c|c|c|c|c|c|c|c|c|}
\cline{1-15}
\multicolumn{1}{|c|}{} & \multicolumn{7}{c|} {QP} & \multicolumn{7}{c|} {MRE}\\
\cline{1-15}
\multicolumn{1}{|c|}{} & \multicolumn{2}{c|}{Rapidity} &\multicolumn{2}{c|}{Temp.}
&\multicolumn{3}{c|}{Mult}
& \multicolumn{2}{c|}{Rapidity} &\multicolumn{2}{c|}{Temp.}
&\multicolumn{3}{c|}{Mult}\\ 
\multicolumn{1}{|c|}{} & \multicolumn{2}{c|}{Y/Y$_p$} &\multicolumn{2}{c|}{(MeV)}
&\multicolumn{3}{c|}{}
& \multicolumn{2}{c|}{Y/Y$_p$} &\multicolumn{2}{c|}{(MeV)}
&\multicolumn{3}{c|}{}\\
\cline{1-15}
b (fm)      & 0-1    & 7-8 & 0-1   &  7-8    & 0-1	& 4-5   & 7-8   & 0-1    &  7-8   & 0-1  
 &  7-8    & 0-1   & 4-5   &  7-8 \\
\cline{1-15}
proton& 0.91 & 0.93 & 8.4 & 5.0   & 0.5 & 1.6 &  1.5& 0.46 & 0.49 & 25.5 & 15.0  & 6.3 &2.6  & 0.3\\
alpha & 0.82 & 0.95 & 14.4&  6.9  & 0.6 & 1.2 &  0.5& 0.46 & 0.71 & 30.1 & 8.4   & 2.9 &0.8  & 0.2\\
\hline
\end{tabular}
\vskip 0.5cm
\caption{Fit parameters for protons and alpha particles in the quasi-projectile (QP) 
and the mid-rapidity component (MRE).}
\end{center}
\end{table}

\begin{figure} 
\begin{center}
\psfig{file=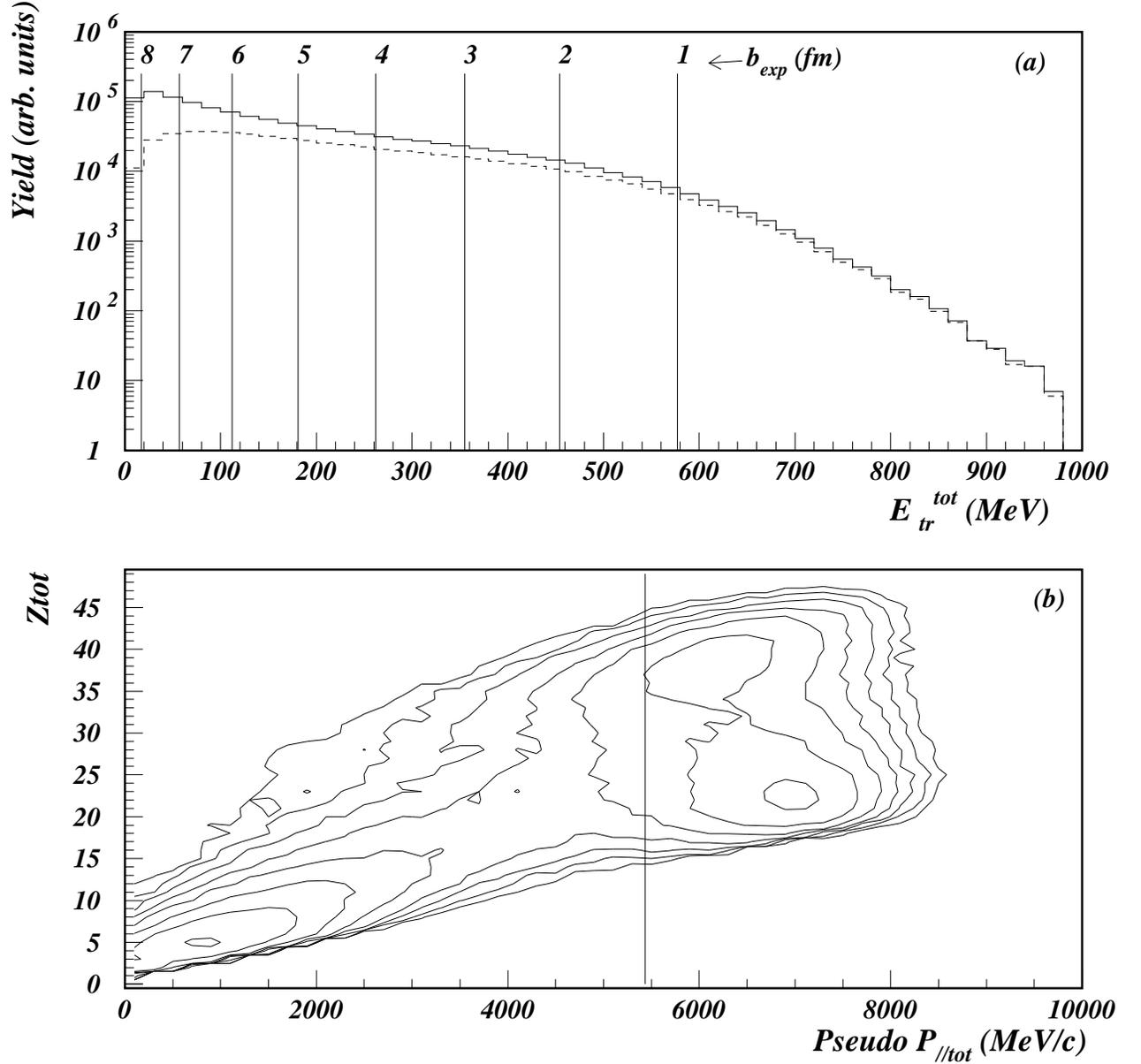,width=18.cm,height=18.cm}
\caption {(a) Total transverse energy distributions for all (full line) and selected 
events (dashed line). 
 Estimated impact parameter values (in fm) corresponding to different $E_{tr}^{tot}$
 intervals are indicated.(b) Total charge versus pseudo total parallel momentum. 
 Only events with a total momentum larger than 
 70 $\%$ of $P_{proj}$ are selected (vertical line). There is a factor two between
 contour levels.}
\end{center}
\label{Fig1}
\end{figure}

\begin{figure}
\begin{center}
\psfig{file=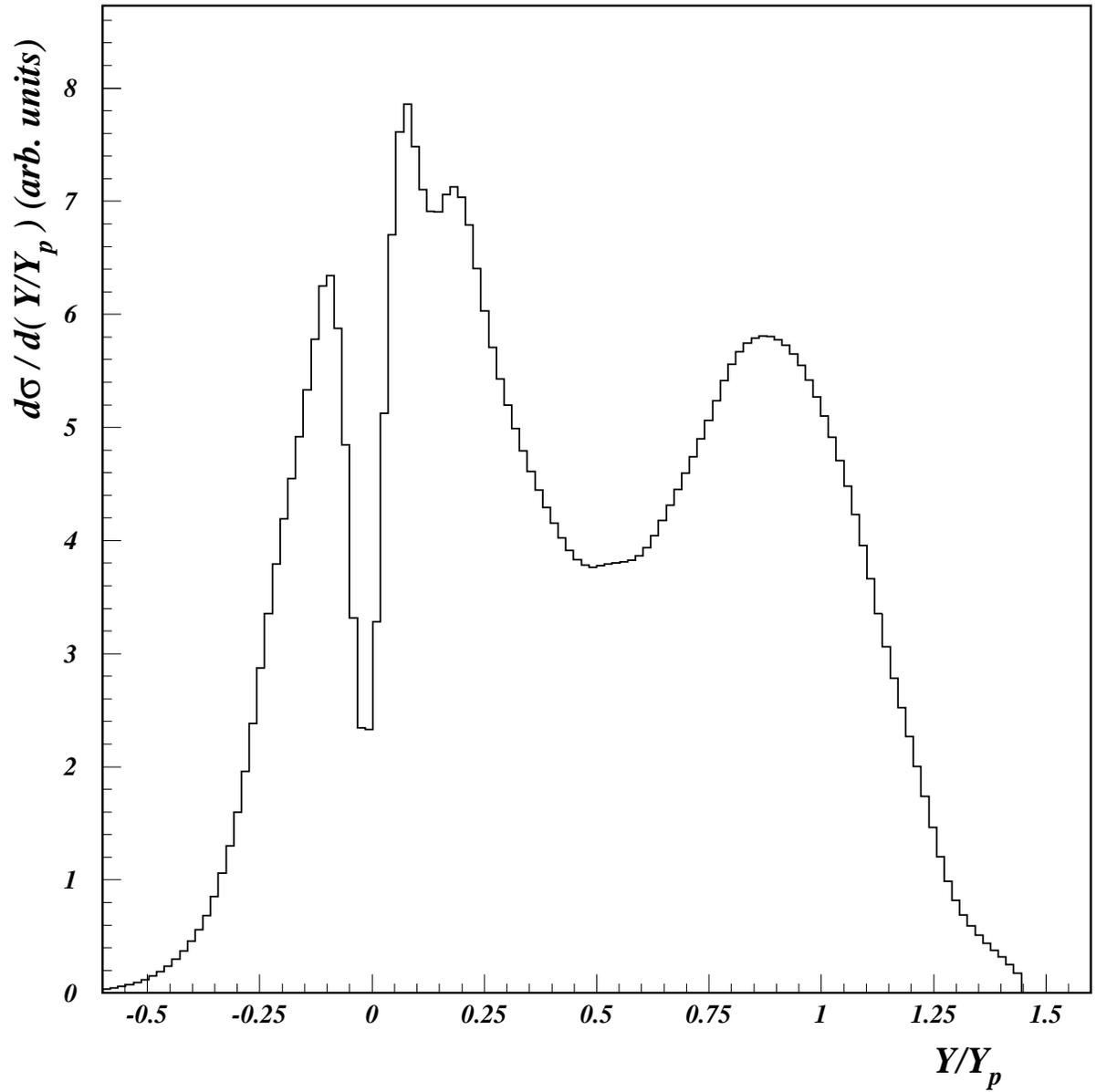,width=18.cm,height=18.cm}
\caption { Proton reduced rapidity distributions at b=6 fm. The hole at $Y/Y_p$=0 is 
due to the target shadow. }
\end{center}
\label{Fig2}
\end{figure}

\begin{figure}
\begin{center}
\psfig{file=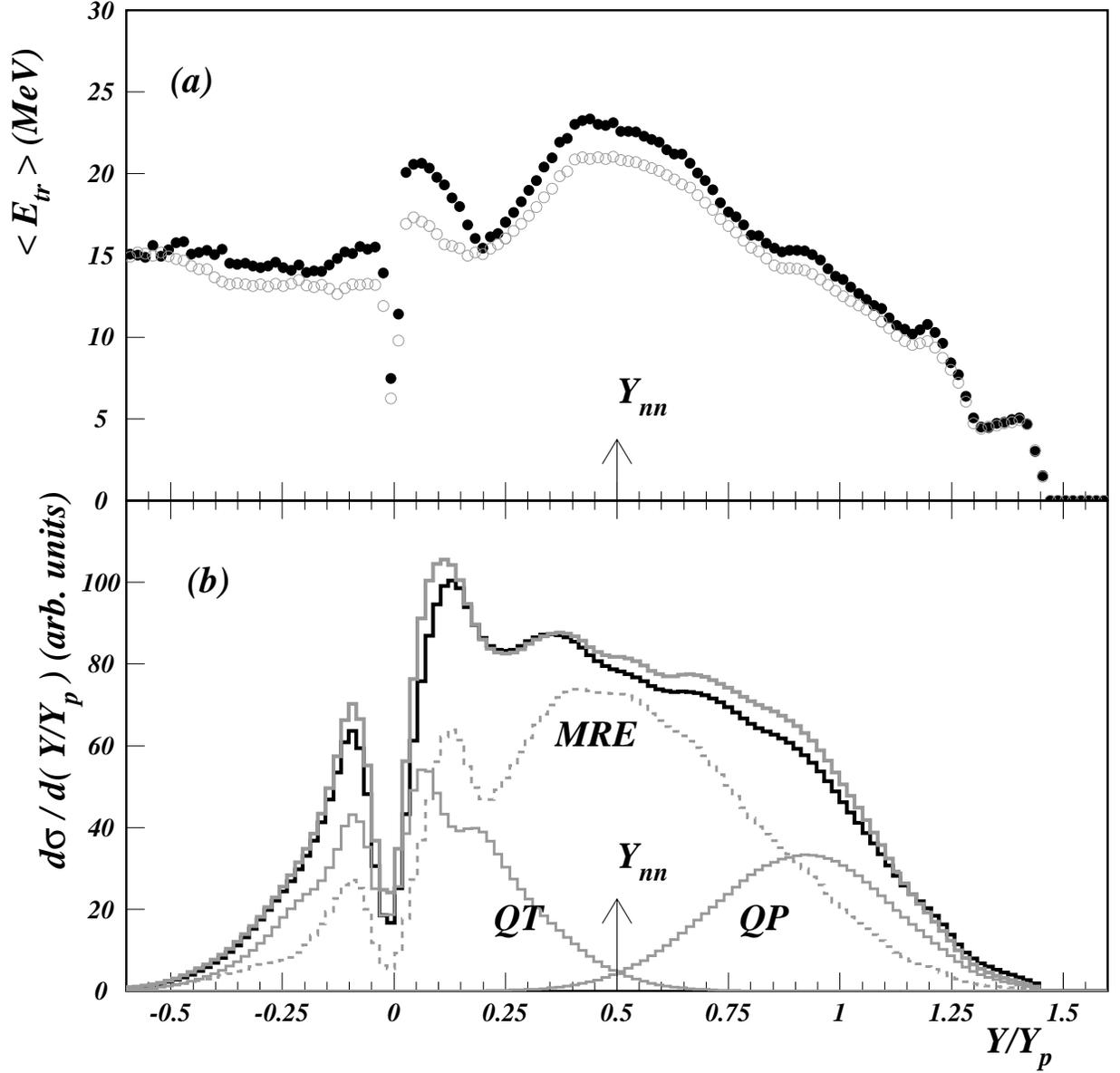,width=18.cm,height=18.cm}
\caption {Data and fit results for $b_{exp}$=3 fm. (a) Average transverse energy $<E_{tr}>$ of 
protons vs reduced rapidity. The experimental data (full circles) are compared 
with the result of a three source fit (open circles). (b) Proton reduced rapidity distribution.
 The experimental data are indicated by the dark histogram and  
fit result by the grey line. The fit contributions of the QP, QT (grey lines) and the
MR (dashed line) are drawn.  The arrow labelled $Y_{nn}$ indicates the
nucleon-nucleon reduced rapidity. For experimental and calculated spectra, 
 an energy threshold of 2 MeV was imposed.}
\end{center}
\label{Fig3}
\end{figure}

\begin{figure}
\begin{center}
\psfig{file=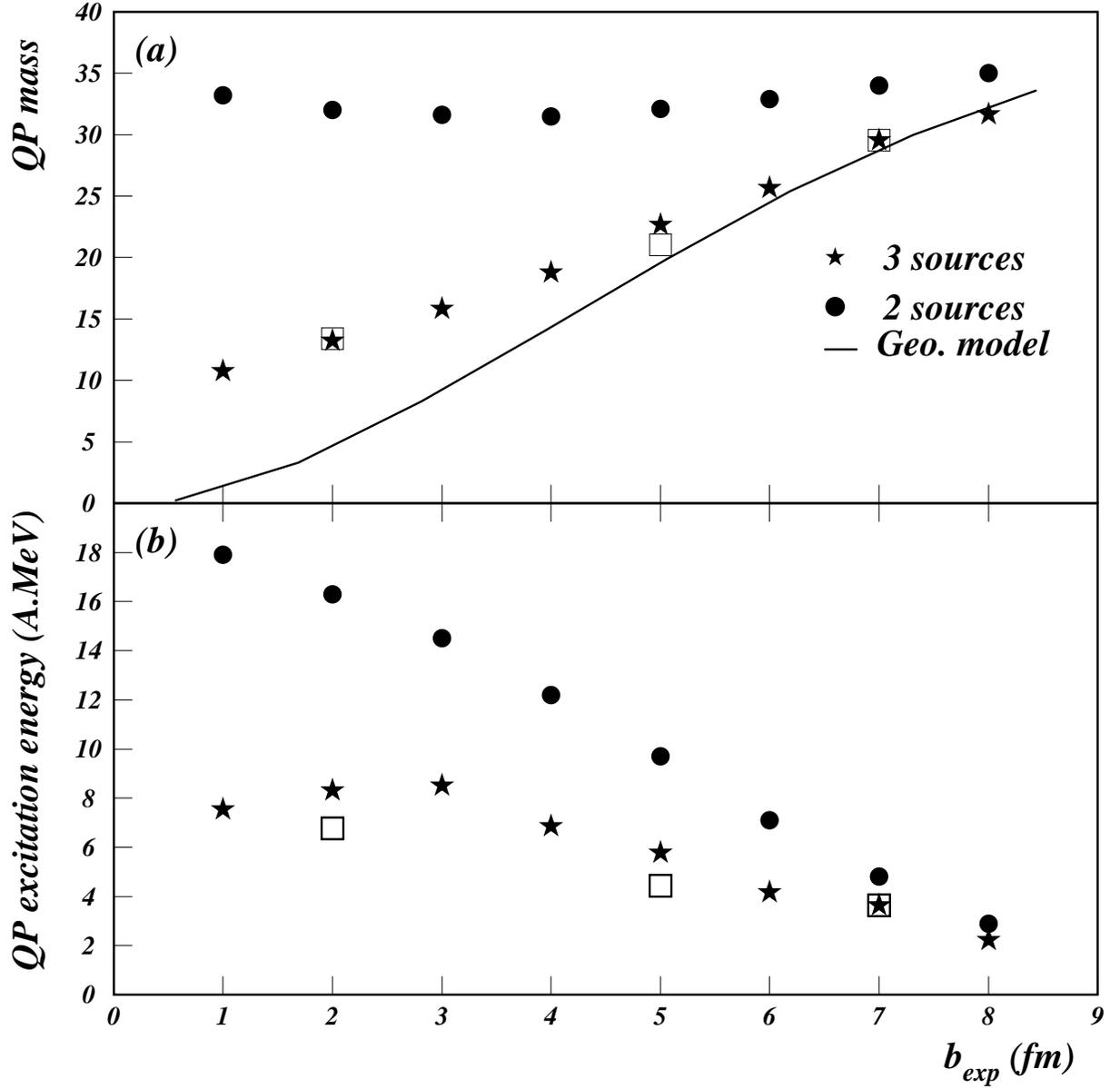,width=18.cm,height=18.cm}
\caption {Average QP mass (a) and excitation energy (b) as a function of the experimental impact
parameter. The line in (a) is the result of a geometrical calculation (see text).
Open symbols show the effect of the different event sortings.}
\end{center}
\label{Fig4}
\end{figure}

\end{document}